\documentclass[twocolumn,showpacs,preprintnumbers,amsmath,amssymb]{revtex4}
\usepackage{graphicx}% Include figure files
\usepackage{dcolumn}% Align table columns on decimal point
\usepackage{bm}% bold math

\begin{document}

\preprint{APS/PRE-SOC}

\title{Markov Chain Hidden behind Power Laws Mechanism of Self-organized Criticality}% Force line breaks with \\

\author{De Tao,~Mao}
\email{detaom@ece.ubc.ca}
% \altaffiliation[Also at ]{Electrical and Computer Engineering Department, UBC University.}%Lines break automatically or can be forced with \\
%\author{, Zhong}%
 %\email{yszdau@tsinghua.edu.cn}
 %\author{Jose.R.~Marti}
%\email{jrms@ece.ubc.ca}
\affiliation{%
Electrical and Computer Engineering Department, University of
British Columbia.
}%

\author{Yisheng, Zhong}
\email{zys-dau@mail.tsinghua.edu.cn}
\affiliation{%
Automation Department, Tsinghua University.
}%

\date{\today}% It is always \today, today,

\begin{abstract}
\hspace*{\parindent}To describe and analyze the dynamics of
Self-Organized Criticality (SOC) systems, a four-state
continuous-time Markov model is proposed in this paper. Different
to computer simulation or numeric experimental approaches commonly
employed for explaining the power law in SOC, in this paper, based
on this Markov model, using E.T.Jayness' Maximum Entropy method,
we have derived a mathematical proof on the power law distribution
for the size of these events.  Both this Makov model and the
mathematical proof on power law present a new angle on the
universality of power law distributions, they also show that the
scale free property exists not necessary only in SOC system, but
in a class of dynamical systems which can be modelled by the
proposed Markov model.
\end{abstract}

%Classification Scheme.
\maketitle

%-*-*-\section{\label{sec:level1}First-level heading:\protect\\ %%\protect\\ The line
%-*-*- break was forced \lowercase{via} \textbackslash\textbackslash}
\section{\label{sec:introduction}Introduction}
\hspace*{\parindent}As a distinctive signature, power law exists in many %research areas related to
real-world systems, such as forest fire \cite{bib:ReedWJ2004}
\cite{bib:HotcomplexityRobustness} \cite{bib:HOT-PL-GCoding},
epidemic
\cite{bib:EpidemicSpreadingInSFNetworks}\cite{bib:HaltingVirusesInSFnetworks},
avalanche, social wealth distribution
\cite{bib:WealthCondensationInParetoMacroEconomies}, blackout of
power systems
\cite{bib:EvidenceSOCinElePowerSystemBlackouts}\cite{bib:AnalysisofElectricPowerDisturbanceData},
and earthquake \cite{bib:pfEarthquakes}
%\cite{bib:WealthCondensationInParetoMacroEconomies}
etc. It is also given various names in these above areas, for
instance, ``Pareto distribution'', ``Zipf's law'',``scale free
distribution'', and ``fat tails'' etc \cite{bib:powerOfDesign}.\\
\hspace*{\parindent} Since the famous work of BTW
(Bak-Tang-Wiesenfeld) \cite{bib:Bak1988}\cite{bib:Bak1988_2},
which was mainly on proving that Self-Organized Criticality (SOC)
was a potential underlying mechanism of any system with the
fingerprint of power
law, %\cite{bib:Bak1995}\cite{bib:BakHowNatureWorks},
there are increasing research efforts on studying the issue of
ubiquitous existence of this scale free property in nature.
However, most results are obtained via comprehensive experiments
of computer simulation, but  the problem why SOC systems and many
other systems are ruled by power law, is
still mathematically unsolved. \\
\hspace*{\parindent}In this study, a Continuous-time Hidden Markov
model is proposed, with which a rigorous proof on the power law
distribution is derived analytically. This study also provides
insights into this novel combination of Continuous-time Markov
model and the Maximum Entropy Principle
\cite{bib:ETJaynessProbabilityTheory}, leading to a better
understanding on the dynamics of these systems with power law
signature.\\
\hspace*{\parindent}The rest of this paper is organized as
follows, Section \ref{sec:related} briefly introduces some
background conceptions, and also a hidden Markov model is
constructed to describe the evolution process of SOC. These
descriptive equations in different states are introduced in
Section \ref{sec:overview Population Model}. The proof on power
law in SOC systems is given in Section \ref{sec:poof of SOC},
simulation results are displayed in Section
\ref{sec:SimulationResults}. Finally, Section \ref{sec:discussion
and conclusion} will give some discussions, conclusions and
possible future work of this study.
%%%%%%%%%%%%%%%%%%%%%%%
\section{\label{sec:related}Background Conceptions and the proposed hidden markov model}
\hspace*{\parindent}This work is closely related to Self-Organized
criticality  and  Continuous-time Hidden Markov Chain theory. A
brief introduction to the basic conceptions and the proposed model
are given here.
%review of the existing work as well as some basic conceptions in these aspects.
\subsection{\label{sec:Overview-SOC}Self-organized Criticality}
\hspace*{\parindent}Self-organized Criticality is a universal
property of systems far from equilibrium, and it is initially
proposed by BTW (Bak-Tang-Wiesenfeld) in
\cite{bib:Bak1988}\cite{bib:Bak1988_2}\cite{bib:Bak1995}\cite{bib:BakHowNatureWorks}.
From then on it has been widely applied to describe the
macroscopic dynamical behavior of an open system near its critical
point, where the system displays a scale invariance feature. But
different to phase transition process in classical physics, at
critical point, it is unnecessary for the SOC system to tune its
parameters to a set of concrete values. And BTW's model
demonstrates that the observed complexity which is emerged in a
robust manner, does not depend on the finely-tuned details of the
system, parameters of the model could be changed widely without
affecting the emergence of critical behavior.
%Besides a high level of nonlinearity,a self-organized
%system has some special properties, like Positive feedback, Negative Feedback and Multiple interactions etc.\\
%\subsection{ model}
%\subsection{Introduction to Statechart}
%\hspace*{\parindent}
%%%--%%\subsection{Introduction on Statechart \cite{bib:Statechart3}\cite{bib:Statechart}\cite{bib:Statechart2}}
%%%--%%\hspace*{\parindent}Statechart is a type of finite state machine,
%%%--%%which can capture the complexity of behavior in dynamic system in
%%%--%%a top-down view. It has extended Moore state machines with many
%%%--%%advanced modelling features, such as concurrency states, composite
%%%--%%states, entering/exit actions, actions on transitions, and guards
%%%--%%etc. These features allow researchers to model more complex
%%%--%%behavior neatly and completely. In this paper, we only use the
%%%--%%simplest features of statechart to model the behaviors of SOC
%%%--%%system. i.e, these transition process that triggered by event
%%%--%%and/or induced by certain preconditions. And the purpose of
%%%--%%applying statechart here is to explore the controllability of SOC
%%%--%%process as well as to represent it clearly.
%features can be quite daunting when you consider how to implement them.
%\subsection{Hidden Markov Model}
%\hspace*{\parindent}\\
\subsection{\label{sec:overview-CTHMM}Continuous-time Hidden Markov Model}
\begin{figure}
\centering
\includegraphics[width=2.50in,height=2.0in,bb=0mm 0mm 199mm
%242mm]{F4statesHMM1.eps}\quad \caption{Four States Continuous-Time
185mm]{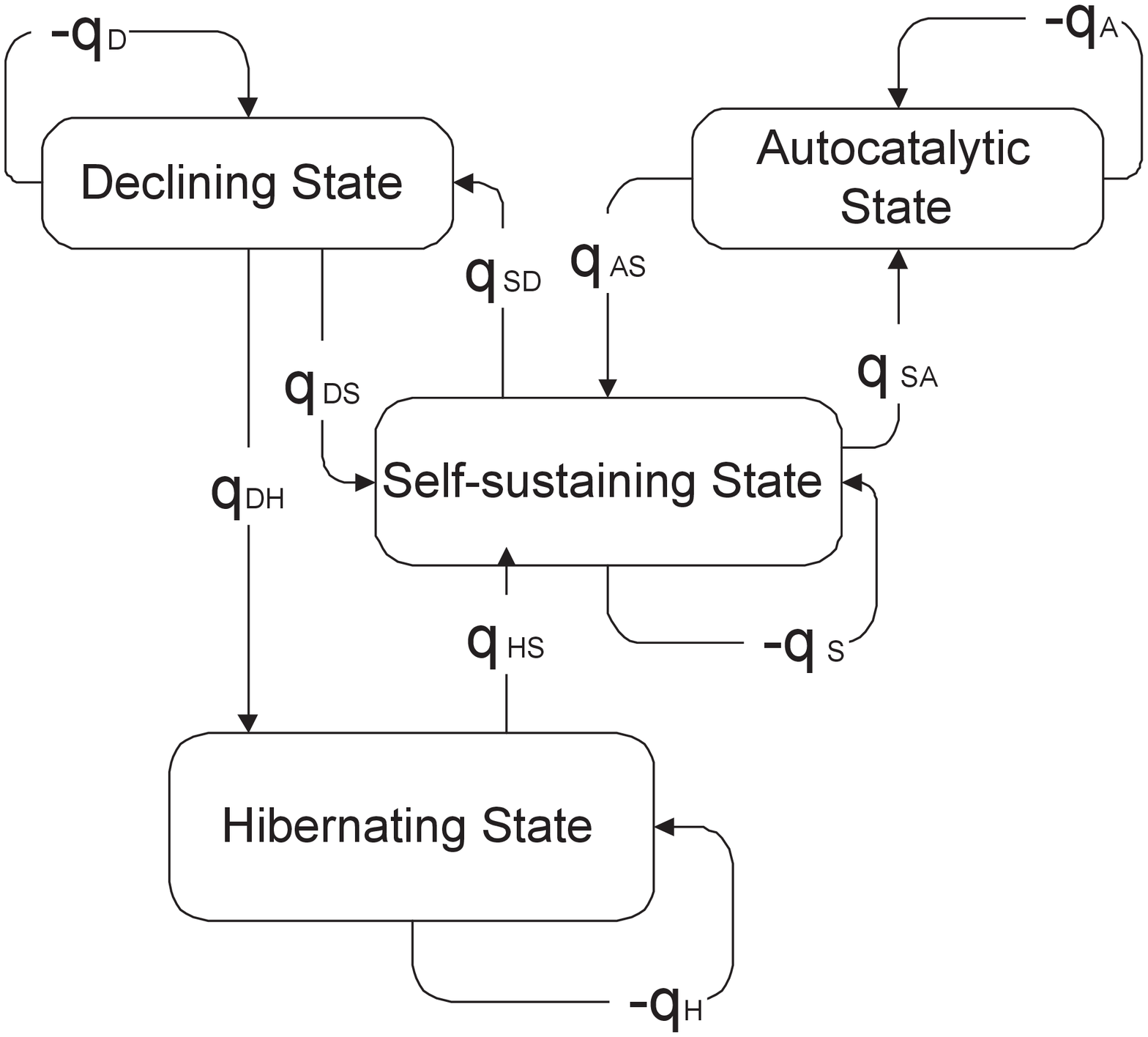}\quad \caption{Four States Continuous-Time
Hidden Markov Model for evolution processes in Self-organized
Criticality.} \label{fig:four_states}
\end{figure}

 \hspace*{\parindent}When observing SOC
phenomena in nature, such as forest fire, earthquake, and
epidemics etc, the most obvious characteristics is their
unexpected emergence. Hence intuitively, their evolution process
can be roughly modelled  into two states: hibernating state and
active state.
%$^{\cite{bib:Statechart}\cite{bib:Statechart2}}$.
However, considering the underlying mechanism for the stochastic
character of SOC, an active state is not sufficient to describe
the complexity of this dynamical property. Thus, the active state
is divided into three sub-states: self-sustaining state, declining
state and autocatalytic state. An intermediate
state--self-sustaining state is introduced here, because for a
continuous-time model, the transition interval between declining
state and autocatalytic state cannot be infinitesimal.\\
\hspace*{\parindent}Suppose for this four-state state machine, the
transition rate between state \emph{i} and state \emph{j} is is
$P(E)_{ij}$, since usually this transition is triggered by events
or induced by a certain type of conditions, thus this
event-triggered or condition-induced transition probability can be
written as,
\begin{eqnarray}
P(E)=P(E_{ex})+P(E_{en}|C)P(C)-P(E_{ex}\cap E_{en}\cap
C)^{\footnotemark[1]} \nonumber
\end{eqnarray}
\footnotetext[1]{In $P(E)$, we only count independent exogenous or
endogenous events, these events that belong to an event cascade
caused by $E_{ex}$ and/or $E_{en}$ are not considered.}
\hspace*{\parindent} Where  $E_{ex}$ denotes an exogenous event
that will trigger a transition from one state to another, $E_{en}$
denotes an endogenous event exists only under a certain condition
$C$, and $P(E_{ex})$, $P(E_{en})$, $P(C)$ are the occurrence
probabilities
of event $E_{ex}$, event $E_{en}$, condition $C$, respectively.\\
\hspace*{\parindent} Assuming these event-triggered and/or
condition-induced transition probabilities $P(E)_{ij}$ are all
statistically stable. %\textbf{Detao: Add some related work here}
%\hspace*{\parindent}%In the above part, the stochastic property
%and other characteristics of SOC process can be phenomenally
%described by a reduced finite state machine in
%Fig.~\ref{fig:four_states}, where we suppose that these
%event-triggered and/or condition-induced transition probabilities
%$P(E)$ are all constant in statistics sense, i.e, the system
%keeps its statistics properties invariant.
Mathematically, these processes can be modelled as a
Continuous-time Markov chain. This four-state Markov model can be
drawn in Fig.~\ref{fig:four_states}. Its transition rates $q_{ij}$
are typically described as the \emph{ij-th} elements of the
Infinitesimal generator matrix \emph{Q}.
In a infinitesimal time interval $\Delta t$, \emph{Q} can be expressed as%$\setcounter{equation}{4}
\begin{eqnarray}
Q=\left(\begin{array}{cccc}
-q_{\mathtt{H}}&q_{\mathtt{HS}}&0&0 \\ %nonumber\\
0&-q_{\mathtt{S}}&q_{\mathtt{SD}}&q_{\mathtt{SA}} \\ %nonumber \\
q_{\mathtt{DH}}&q_{\mathtt{DS}}&-q_{\mathtt{D}}&0 \\ %\nonumber \\
0&q_{\mathtt{AS}}&0&-q_{\mathtt{A}} %\eqno(2)
\end{array}\right)
\end{eqnarray}
where $q_{ij}$ is the transition rate between state
\emph{i} and state \emph{j}, $q_{i}$ is the self-transition rate of state \emph{i}.%Here,the diagonal element $q_{ii}$ is given by
%$q_{ii}=-q_{i}=-\sum_{j\ne i}q_{ij}$.\\
\\
\hspace*{\parindent} \hspace*{\parindent}When time approaches
infinity, the stationary distribution of the markov states
$\pi=(\pi_{\mathtt{H}},\pi_{\mathtt{S}},\pi_{\mathtt{D}},\pi_{\mathtt{A}})
$ can be calculated by
\begin{eqnarray}
\pi=e\cdot
(Q+E)^{-1}=(\pi_{\mathtt{H}},\pi_{\mathtt{S}},\pi_{\mathtt{D}},\pi_{\mathtt{A}})
\end{eqnarray}
\\Where
\begin{eqnarray}
E=\left(\begin{array}{ccc}
1&\cdots&1 \nonumber\\
%1&1&1&1 \nonumber \\
%1&1&1&1 \nonumber \\
&\cdots& \nonumber \\
%\multicolumn{2}{c}{\dotfill}
1&\cdots&1 \nonumber
\end{array}\right)_{4\times4},\nonumber \\
e=[1,1,1,1],\nonumber \;
\pi_{\mathtt{H}}+\pi_{\mathtt{S}}+\pi_{\mathtt{D}}+\pi_{\mathtt{A}}=1.\nonumber
\end{eqnarray}
and $\pi_{\mathtt{H}}$,$\pi_{\mathtt{S}},\pi_{\mathtt{D}},
\pi_{\mathtt{A}}$ denotes the stationary distribution of
Hibernating state, Self-sustaining state,Declining state, and Autocatalytic state, repectively.\\
%%%\hspace*{\parindent}Note that these parameters in the
%%%infinitesimal generator matrix $Q$ of this proposed Markov model
%%%can not be directly extracted from the observed data, and these
%%%three sub-states cannot be discriminated from each other. as one
%%%type of finite state machine, this statechart model can also be
%%%regarded as a certain type of markov
%%%model, and from these reasons above, the statechart model should be categorized as \emph{Hidden Markov Model}.\\

%(a complicated detail can be expressed
%in the Section \ref{sec:discussion and conclusion}. with the
%mechanism of Boltzman equation.)}\\%
%here you can give a figure to describe  it
%what is the hypothesis of this paper?
%is the system heterogenous or homogenous or purely uniform?
\section{\label{sec:overview Population Model}The evolution equations in different
states} \hspace*{\parindent}For a SOC system far from equilibrium,
when it is approaching its critical point, instead of entering a
state of phase transition entirely, parts of its unit population
will evolve into a meta-stable state.
%% Suppose that there
%%is a distribution to describe the stable degree of each individual
%%unit in the system, a partial differential equation(PDE) can be
%% introduced here to model these units with various stable degrees.
%% If the entire population $p(t)$ is below a certain threshold $p_{min}$, the whole
%%system will stay in the hibernating state. Also we suppose that
%% the meta-stable units has a distribution, $p(x_i,t)$, where $x_i$
%% can be regarded as the ``activation energy''
%%  for describing the unstable degree of a unit, and $t$
%% denotes the system evolution time. Given $x_{i0}$ is a certain
%%threshold value for activation energy, if $||x_i||>x_{i0}$, the
%%PDE model for meta-stable units in the population can be derived
%%as follows. \setlength{\arraycolsep}{2.5pt}
%%\begin{eqnarray} \label{eq:population}
%%\frac{\partial p(x_i,t)}{\partial t}+\frac{\partial p(x_i,t)}{\partial x_i}=-d(x_i,t)\cdot p(x_i,t) \nonumber\\
%%t=0:p=p_{min}(x_i)( x_i \geq 0)& \nonumber \\
%%x_i=0:p(0,t)=\int_{a}^{A}b(\xi)\cdot p(\xi,t)d\xi(t\ge0)
%%\end{eqnarray}
%\right.\]
%%In Eqn.(\ref{eq:population}), $a$ is a threshold value; $A$ is the
%%maximum unstable degree value for $x$; $d(x,t)$ is the death rate
%%and $b(x,t)$ is the birth rate; and $p_{min}(x,t)$ is the minimum
%%size of event that can be observed in the SOC process. \\

\hspace*{\parindent}To demonstrate the evolution process of SOC
systems, we can analyze its activities in autocatalytic state
firstly. In an autocatalytic state, as most of these clustered
meta-stable units are in critical state, a small event can trigger
a large disaster that propagates quickly as well as greatly changes
the stable-degree distribution of these units.\\
\hspace*{\parindent}For an ongoing event in autocatalytic state,
let $\xi=\frac{d\vec{n}}{dt}$ be the unfolding speed at the normal
direction $\vec{n}$ of this event, via a first-ordered linear
approximation, $\xi=\frac{d\vec{n}}{dt}$ can be determined in
proportion to the divergence ($\nabla\cdot\Phi$)of the energy flux
$\Phi$ dissipating across its boundary.
%%\emph{Additionally, it can also be interpreted as the integral total
%%effect of the event to its boundary units?}.%, as illustrated in Fig.~\ref{fig:collective_action}.
For instance, in BTW's sandpile model for SOC, during an event,
these triggered meta-stable units will release their energy to
these peripheral units, among which these meta-stable ones will be
triggered to release their energy, and these stable one will
absorb the energy and become unstable. The equations to describe
 the above processes is derived as,
\begin{eqnarray}
\xi=\frac{d\vec{n}}{dt}=k_1 \nabla\cdot \Phi \nonumber \\
\nabla\cdot \Phi =\frac{\oint d \Phi}{\oint d \l} \nonumber \\
\oint d \l\cdot d \vec{n} = d A
\end{eqnarray}
As the total energy flux $\Phi$ produced by the event is in
proportion to its current size $A$, thus
\begin{eqnarray}
\oint d \Phi= k_2\iint\limits_{A} d\sigma=k_2\cdot A
\end{eqnarray}
\hspace*{\parindent}Where $k_1$ and $k_2$ denote these linear
coefficients, $d\sigma$ is the differential of area $A$. %constant quantities.
The final expression of the above equations can be converted to:
\begin{eqnarray}
\frac{dA}{dt}=k_1\cdot k_2
A=\overset{+}{\mathbf{r}}A\;(\overset{+}{\mathbf{r}}>0)
\end{eqnarray}\label{eqn:autocatalytic}
As observed in the above equation, the coefficient
$\overset{+}{\mathbf{r}}$ stands for the intensity of
positive-feedback effect in autocatalytic state, which is
determined by the dependency range or correlated degree among the
different units in the system.\\
%As in the figure By only considering the birth or death process,
%as the product of the linear $k_1 k_2 $ as , to
%\begin{figure}
%\centering
%\includegraphics[width=2.0in,height=2.0in,bb=47mm 154mm 146mm
%255mm]{forestfire1.eps}\quad \caption{During the autocatalytic
%state, the unfolding speed of an event is determined by the
%density of the mean
% dissipative energy flux across its boundary.}
% \label{fig:collective_action}
%\end{figure}\\
\hspace*{\parindent} Accompanying an ongoing event, a similar
analysis \emph{ process} can also be applied on its declining
states, for an event cannot trigger these meta-stable units to
make them release energy all the time, at micro time scale, these
units at the event's boundary may absorb these released energy,
making these meta-stable units become stable and the possible
large event will shrink to a small size, this can be seen as an
reversible process of the autocatalytic state, similar to the
above analysis, we can get an equation to describe the declining
state,
\begin{eqnarray}
\frac{dA}{dt}=\overset{-}{\mathbf{r}}A
\;(\overset{-}{\mathbf{r}}<0)
\end{eqnarray}\label{eqn:decliningState}
Similarly, the coefficient $\overset{-}{\mathbf{r}}$ stands for
the intensity of negative-feedback effect in declining state,
which is also determined by the dependency range or correlated
degree among the different units in the system.\\
\hspace*{\parindent}As in hibernating state there is no event,
while in self-sustaining state, there is an ongoing event in the
system. It is obvious that in self-sustaining state, the event's
size keeps invariant, therefore the evolution time in this state
can be omitted\emph{ in our model}, since  only the declining
state and the autocatalytic state are essential in determining the
final size of the event. An evolution equation for the size of the
event can be proposed as follows:
\begin{eqnarray}
\frac{dX(t)}{dt}=[\overset{+}{\mathbf{r}}\cdot \delta(S_{t}-S_{A})+\overset{-}{\mathbf{r}}\cdot \delta(S_{t}-S_{D})]X(t) \nonumber\\
%=(\overset{+}{\mathbf{r}}\delta(S_{t}-S_{A})+\overset{-}{\mathbf{r}}\delta(S_{t}-S_{D}))p(t) \nonumber\\
t=0:X=X_{0}.\label{eq:master_equation}
\end{eqnarray}
% \right.%\]
\hspace*{\parindent}In Eqn.(\ref{eq:master_equation}), $X(t)$ is
the size of the event at time $t$, $\delta$ denotes Dirac
function, $S_{t}$ means the system's state at time $t$, $S_{_A}$
($S_{_D}$) denotes the system's autocatalytic (declining) state at
time $t$; $\overset{+}{\mathbf{r}}$ ($\overset{-}{\mathbf{r}}$)
corresponds to the expanding (shrinking) speed of meta-stable
units in an autocatalytic (declining) state.\\%$S_{_A}$($S_{_D}$).\\\\
\hspace*{\parindent}According to Eqn.(\ref{eq:master_equation})
above, an event's final size can be expressed as
\begin{eqnarray}
X(t)=X_{0}\cdot  e^{\overset{+}{\mathbf{r}}  {\sum_{i}
\mathbf{t}^{\mathtt{_A}}_{\mathtt{_i}}} + \overset{-}{\mathbf{r}}
{\sum_{j}
\mathbf{t}^{\mathtt{_D}}_{\mathtt{_j}}}} \label{eqn:master_equation solution}
%\nonumber\\t=0:X=X_{0};\overset{-}{\mathbf{r}}<0;\overset{+}{\mathbf{r}}>0.
\end{eqnarray}
\hspace*{\parindent}Where $\sum_{i}
\mathbf{t}^{\mathtt{_A}}_{\mathtt{_i}}$ ($\sum_{j}
\mathbf{t}^{\mathtt{_D}}_{\mathtt{_j}}$) represents the
accumulated duration that the event has stayed at the
autocatalytic (declining) state. $i$ ($j$) represents that how
many times that the event has visited the autocatalytic
(declining) state.

\section{\label{sec:poof of SOC}SOC as a stationary process in the view of the law of large numbers}
% ( or ergodicity)}
\subsection{Stable Interaction Hypothesis}
\hspace*{\parindent}In SOC process, such as forest-fire model and
BTW's sandpile model, let $\mathcal{S}_s(t)$,$\mathcal{S}_a(t)$
and $\mathcal{S}_d(t)$ denote as  the ``negative
entropy"$^{\footnotemark[2]}$ stored in system, absorbed from the
surrounding environment, and dissipated into the environment,
respectively. Via the laws of conservation, there exists the
following relation\footnotetext[2]{The ``negative entropy'' here
can be defined in a general way, including mass \emph{M}, entropy
\emph{S}, information \emph{I}, energy \emph{E} and other
combinational
forms of them.},% among them:
\begin{eqnarray}
\mathcal{S}_d(t)=\mathcal{S}_a(t)-
\mathcal{S}_s(t)\label{eqn:conservation relationship}
\end{eqnarray}
Which means in a short term, $\mathcal{S}_d(t)$ is temporally
determined
by $\mathcal{S}_a(t)$ and $\mathcal{S}_s(t)$.\\
\hspace*{\parindent}However, in the long term, given the system's
environments keep stable,i.e., $\mathcal{S}_a(t)$ and
$\mathcal{S}_d(t)$ are all stationary processes, as for any SOC
system under study,  the negative entropy stored in it
$\mathcal{S}_s(t)$ can be deemed as a finite quantity. Thus when time goes infinity,%refore
%\begin{eqnarray}
%\lim_{_{t \to
%\infty}}\frac{1}{t}(\mathcal{S}_a(t)-\mathcal{S}_d(t))=\lim_{_{t
%\to
%\infty}}\frac{1}{t}\mathcal{S}_s(t)=0 \nonumber\\
%\end{eqnarray}\\
%expressed in integral form, it becomes
\begin{eqnarray}
\lim_{_{(t-t_0) \to \infty}}\frac{\int_{t_0}^t
\mathcal{S}_a(\tau)d\tau}{t-t_0}=\lim_{_{(t-t_0) \to \infty}}
\frac{\int_{t_0}^t\mathcal{S}_d(\tau)d\tau}{t-t_0}  \nonumber \\  %(t\to+\infty)
%\Rightarrow\overline{\mathcal{S}_a(t)}=\overline{\mathcal{S}_d(t)}
\label{eqn:Sa_Sd_IO}
\end{eqnarray}
 %the total energy absorbed from and dissipated into the environment can be taken as equal:
 which means %wherre $\overline{\mathcal{S}_a(t)}$ ( $\overline{\mathcal{S}_d(t)}$ )
 %denotes the average negative entropy absorbed from (dissipated into) environment.
  %Therefore,
  in long term, the entropy absorbed
from the environment and
dissipated into it can be deemed as equal.\\

\hspace*{\parindent}For these processes with  property of
statistical stationary, determined by Eqn.(\ref{eqn:conservation
relationship})and Eqn.(\ref{eqn:Sa_Sd_IO}), the average quantity
of negative entropy exchanged between the system and its
environment $\Delta S_{ex}$ keeps at an invariant level. Given any
length-fixed time window $[\mathtt{t_i},\mathtt{t_{i}+\Delta T}]$,
the  number of events $n_i$ within it may fluctuate from time to
time. But in a long term, for $n$ times of observations, according
to the law of large numbers,
 the average number of observable
events $\overline{n}=\frac{\sum_i n_i}{n}$ within this time window
 $\Delta T$ can be regarded invariant, therefore the
average time interval  between
events $T_0=\frac{\Delta T}{\overline{n}}$ should also be an invariant quantity.\\
%Based on the above analysis, we suppose
%that the average interval of $N$ events occurred during time $t$
%in SOC process is $\mathtt{T_0}$,
 %therefore
%\begin{eqnarray}
%\mathtt{T_0^{'}}=(\mathtt{T_1}+\cdots+\mathtt{T_{N}})/N=t/N \nonumber\\
%\lim_{_{N\to+\infty}}P(|\mathtt{T_0^{'}}-\mathtt{T_0}|<\varepsilon)= 1
%\end{eqnarray}
\hspace*{\parindent}As shown in above sections, SOC process has
been modelled by a continuous-time Hidden Markov Chain. During an
event, the average time  in its autocatalytic state (declining
state) is $\pi_{\mathtt{A}}\mathtt{T_0}$
($\pi_{\mathtt{D}}\mathtt{T_0}$), and from the properties of
continuous-time Hidden Markov model, obviously, $\mathtt{T_0}$,
$\pi_{\mathtt{A}}$ and $\pi_{\mathtt{D}}$ are all invariant
quantities.
\subsection{Proof of Power Law Distribution in SOC}
\hspace*{\parindent}Suppose $x_m$ denote the minimum size of an
observable event in the SOC system under study, for instance, in
BTW's sandpile model, $x_m$ might represent a sand grain, and in
forest fire model, $x_m$ might represent a tree. Assuming that all
the observable events in the system have the same initial size
$x_{m}$, by the solution of Eqn.(\ref{eqn:master_equation
solution}), it is obvious that only when
$(\overset{+}{\mathbf{r}}\pi_\mathtt{A}+\overset{-}{\mathbf{r}}\pi_\mathtt{D})\ge
0$, the event is observable.\\
\hspace*{\parindent}During its evolution process, assuming
\emph{N} events with different size $X_i\; (X_i\geq x_m,
i=1,\cdots,N)$ can be observed in the SOC system. Suppose they are
ruled by a distribution $f(x)$, to histogram them into a double
logarithm coordinates via their sizes, these events can be
quantized into \emph{m} different sets or intervals, for instance,
in set $l$ $(l\in \{1,\cdots, m\})$, there are
$N_{l}\;(\sum_{l=1}^m N_{l}=N)$ events, and in this set, the size
difference  between
 event $X_{l}^j$ and event $X_{l}^k$ should be less than $\delta_{X_{l}}$.
 Here $\delta_{X_l}$ is the threshold
value for set $l$, ensuring in it $\prod_{j=1}^{N_l}(X_l^j)$
 is almost equal to $(X_l)^{N_l}$ .\\
\hspace*{\parindent} Let define quantity \emph{L} as
 \begin{eqnarray}
L\, \stackrel{\mathrm{def}}{=}\, \prod_{i=1}^{N} X_i
=\prod_{l=1}^m(X_l)^{N_l}
\end{eqnarray}
\hspace*{\parindent}Based on Eqn.(\ref{eqn:master_equation
solution}), \emph{L} can be expressed as,
\begin{eqnarray}
 L
=x_{m}^{_N}\; e^{\sum (\overset{+}{\mathbf{r}}
\mathbf{t}^{\mathtt{_A}}_{\mathtt{_{l,i}}}
+\overset{-}{\mathbf{r}} \mathbf{t}^{\mathtt{_D}}_{\mathtt{_{l,i}}})N_l} %(\sum_{i=1}^nN_i=N)
\label{eq:L}
\end{eqnarray}
\hspace*{\parindent} In Eqn.(\ref{eq:L}),
$\mathbf{t}^{\mathtt{A}}_{l,i}$ ($\mathbf{t}^{\mathtt{D}}_{l,i}$)
means the duration that the system stays at the autocatalytic
 (declining) state in the event \emph{i} of set \emph{l}.\\
\hspace*{\parindent}Let define quantity $\psi$ as
\begin{eqnarray}
\psi \, \stackrel{\mathrm{def}}{=}\,\frac{lnL}{N} =ln(x_{m}
e^{{{\sum
(\overset{+}{\mathbf{r}}\mathbf{t}^{\mathtt{A}}_{\mathtt{_{l,i}}}
+\overset{-}{\mathbf{r}}\mathbf{t}^{\mathtt{D}}_{\mathtt{_{l,i}}})}}\frac{N_l}{N}})
\end{eqnarray}
\hspace*{\parindent} It can also be expressed as
\begin{eqnarray}
\psi=\frac{ln(\prod(X_l)^{N_l})}{N} =\sum \frac{N_l}{N}\; ln(X_l)
\label{psi_2}%=ln{x_m}+\frac{\sum (T_{_{Ai}}-T_{_{Di}} )}{N}.%........???? and path integral
\end{eqnarray}
As
 \begin{eqnarray}
 \lim_{_{\delta_{X_l} \to 0}}\frac{N_l}{N} \to f(X_l)
\end{eqnarray}
as \emph{f(x)} denotes  as the events' probability density
function. Eqn. (\ref{psi_2}) can be expressed as
\begin{eqnarray}
\psi=\int_{x_{m}}^{\infty} f(x) lnx\; dx \label{psi2}
\end{eqnarray}
\hspace*{\parindent} therefore $\psi$ is equal to the expectation
of \emph{lnx}, as the stationary distribution of this
continuous-time Markov chain is
$\pi=(\pi_\mathtt{H},\pi_\mathtt{S},\pi_\mathtt{D},\pi_\mathtt{A})$,
and also because
\begin{eqnarray}
\psi=E\{lnx\}=ln(x_{m})+(\overset{+}{\mathbf{r}}\cdot
\pi_\mathtt{A}+\overset{-}{\mathbf{r}}\cdot \pi_\mathtt{D})\cdot
\mathtt{T_0} \label{Eqn:alpha-expression}
\end{eqnarray}
as this SOC process has been assumed as a statistical stationary
process, $x_m$, $\overset{+}{\mathbf{r}}$,
$\overset{-}{\mathbf{r}}$, $\pi_\mathtt{D}$, $\pi_\mathtt{A}$ and
$\mathtt{T_0}$ are all constants, $\psi$ can be deemed as an invariant quantity. \\
\hspace*{\parindent}The entropy of the system can be expressed as
\begin{eqnarray}J(f)=-\int_{x_{m}}^{\infty} f(x) ln f(x)\; dx
\end{eqnarray}%which actually is determined by the event's probability density function $f$,

according to Eqn.(\ref{Eqn:alpha-expression}) and normalization
condition, via lagrangian
multiplier method, %we can add (\ref{normalization})(\ref{psi2})each
% multiplied with a lagrangian multiplier factor,and get
the above equation can be converted in the following form:
\begin{eqnarray} J(f)=-\int_{x_{m}}^{\infty} f(x) lnf(x)\; dx
+\lambda_1 (\int_{x_{m}}^{\infty} f(x)\;  dx -1)\nonumber\\
+ \lambda_2 (\int_{x_{m}}^{\infty} f(x) ln x\; dx-\psi)\;\;\;
\end{eqnarray}
 \hspace*{\parindent}By E.T.Jaynes' Maximum Entropy
 Principle \cite{bib:ETJaynes1957I}, we can get
\begin{eqnarray}
f(x)=e^{\lambda_1}x^{\lambda_2}\nonumber\\
%C_1=-(\lambda_2+1)x_{m}^{-(\lambda_2+1)}(\lambda_2+1<0)
\end{eqnarray}
let $\alpha=-(\lambda_2+1)$, we can get
%according to the normalization condition (\ref{normalization}),
%\[\]therefore
\begin{eqnarray}
f(x)=\alpha x_{m}^{\alpha} x^{-\alpha-1} \;(\alpha>0)
\label{alpha}
\end{eqnarray}
%\hspace*{\parindent}where $\alpha$ is the slope index.\\
 \hspace*{\parindent}As
\begin{eqnarray}
E\{lnx\}=\int_{x_{m}}^{\infty}\alpha x_{m}^{\alpha}x^{-\alpha-1}
lnx \;dx=lnx_{m}+\frac{1}{\alpha} \label{expectationLnx}
\end{eqnarray}
%as \begin{eqnarray} E\{lnx\}=ln E\{x\} \end{eqnarray}
 from Eqn.(\ref{Eqn:alpha-expression}) and Eqn.(\ref{expectationLnx}),
 %here we define $\zeta=\alpha ^{_{-1}}$ as
%the unfolding index, which can be expressed as
\begin{eqnarray}
\alpha=[(\overset{+}{\mathbf{r}}\pi_{\mathtt{A}}+\overset{-}{\mathbf{r}}\pi_{\mathtt{D}})\mathtt{T_{0}}]^{_{-1}}
\end{eqnarray}
%\hspace*{\parindent}Note in Fig.~\ref{fig:four_states},
%considering an irreversible situation, corresponding to the
%autocatalytic state, the declining state as an vertible process
%can be omitted, thus the four-state Markov model that illustrate
%general case can be reduced into a three-state one, as illustrated
%in Fig.~\ref{fig:threestateschart}, where
%\[\zeta=\alpha^{_{-1}}=\overset{+}{\mathbf{r}}\pi_{\mathtt{A}}\mathtt{T_{0}}\]
%\begin{figure}
%\centering
%\includegraphics[width=2.50in,height=2.250in,bb=47mm 140mm 170mm
%257mm]{threestatesChart4ForestFire.eps}\quad \caption {The three
%states chart model for ideal Forest Fire Model. Since its
%evolution or unfolding process is not a reversible one, the
%declining state in the four-state model is omitted.}
 %\label{fig:threestateschart}
%\end{figure}\\
%\section{analysis on computer simulation results and real-life SOCsystem's recorded data}
%\hspace*{\parindent}descretize the Statechart-based CTHMM;simulation;Histogram of power law;\\
%\hspace*{\parindent}real-life data; parameter-estimation;Histogram
%of power law;\\
%\hspace*{\parindent}Conclusion; \subsection{Parameter estimation
%and prediction of Hidden Markov Model for SOC system}
%\hspace*{\parindent}equations,\\
%dynamic programming,\\
%prediction model;\\

\section{\label{sec:SimulationResults}Simulation Results}
\begin{figure}
\centering
\includegraphics[width=2.50in,height=2.50in,bb=23mm 85mm 204mm
240mm]{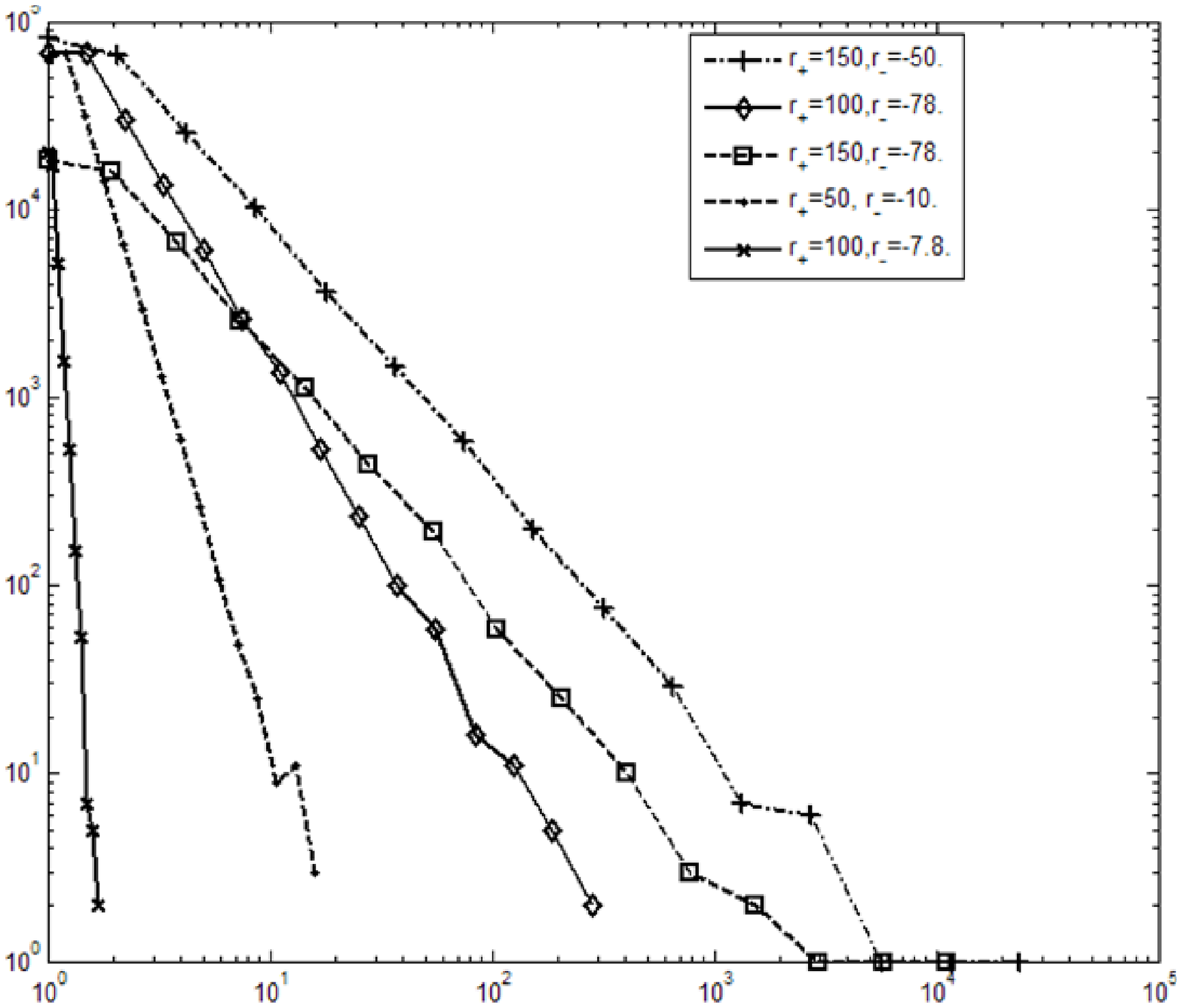}\quad \caption {Power Law distributions with
different simulation parameters. where $\mathbf{r}_{\pm}$ is equal
to $\overset{\pm}{\mathbf{r}}$ noted in the above sections.}
 \label{fig:power law simulation Results}
\end{figure}

\hspace*{\parindent}Since the average periodicity or interval
$T_0$ between events is a constant quantity, and the distribution
for the holding times
 of this four-state Markov model $\{\pi_A,\pi_D,\pi_h,\pi_S\}$  is
stationary, using the discrete time method introduced in
\cite{bib:DTMethods4SimCTMC}, we get different power law
distribution by simulation at different parameters
$\overset{\pm}{\mathbf{r}}$,
 as shown in Fig. \ref{fig:power law simulation Results}. \\

\section{\label{sec:discussion and conclusion}Conclusion and future work}
%In a more general sense, these events are not necessary fully
%developed, they can be these ongoing events that observed
% or killed randomly.\\

\hspace*{\parindent}The SOC system under study and the
environments are all supposed to be statistically stationary, this
can be correct for almost all cases in real-world and artificial
systems. Though observed in a short term, these parameters, like
transition rates, are usually not as static as we expected, which
means  flux of energy, mass, information and their combinational
forms are not constant quantities, the frequency of event will
fluctuate in this short term.\\
\hspace*{\parindent}This continuous-time Hidden Markov Model can
give us  a framework to describe the dynamical properties  not
only  for these SOC systems, but for these regular systems
different to SOC, yet owning positive and/or negative feed-back
states during their evolution processes.
%It offers us a potential way to unify these systems of very
%different origins, no matter they are SOC systems or not, and no
%matter they are revertible or not.
The ubiquity of power laws is mainly because the ubiquity of these
systems that can be described by this Markov model.\\
%\section{\label{sec:discussion and conclusion} future work}
%\subsection{Prediction of the events based on Hidden Markov Chain}
\hspace*{\parindent}One  potential direction based on this model
is to unify these stochastic processes with power law fingerprint,
such as Geometric Brownian motion, Discrete multiplicative
process, Homogeneous birth-and-death process,
 and Galton-Walton branching process etc in \cite{bib:ReedWJ2002}, Forrest-fire model
 in \cite{bib:SOC_FFmodel}, and Yule process
 in \cite{bib:NewmansPLParetoDistributionandZipfsLaw} etc.\\
\hspace*{\parindent}As human behavior can affect the bidirectional
flux between the systems and its environments, there is
possibility on controlling the events frequency and their size
distribution, so as to reduce the probability of large scale
events and prevent the collapse of the whole systems. Some
applications and academic work based on Highly Optimized Tolerance
(HOT) have been done in
\cite{bib:HotcomplexityRobustness}\cite{bib:HOT-PL-GCoding}
\cite{bib:HOT-DynamicsandChangingEvironments}
\cite{bib:Mutation-HOT}\cite{bib:JohndoyleHOT} etc, but as our
model has different features from HOT, a lot of research studies
are still very necessary at this direction.

%\begin{acknowledgments}
%We wish to acknowledge the support of the author community in
%using REV\TeX{}, offering suggestions and encouragement, testing
%new versions, \dots.
%\end{acknowledgments}
\appendix
%\section{Appendixes}
\section{Life time of an event and its multi-time scale property}
%\subsection{the definition of event in the systems of SOC}
%what is an event? \\
\hspace*{\parindent}When the state of a SOC system is in its
hibernating state, there is no event in the system, otherwise when
an event happens, by passing through the self-sustaining state,
the system will escape its hibernating state into any one of these
three active states -- self-sustaining, declining and
autocatalytic states. If the event keeps on, the system state will
jump among the active states stochastically. Therefore in this
Markov model, the lifetime of an event is the duration since  its
last escape from the hibernating state until the next return to
it.\\
\hspace*{\parindent}The multi-time scale property of this
continuous-time Markov Chain model is displayed in the following
Fig.\ref{fig:Multi-time scale model},where A(D,S,H) stands for the
four states of the Markov model respectively. The framed section
is zoomed out to show the detail information of this Multi-time
scale model.
\begin{figure}
\centering
\includegraphics[width=2.50in,height=2.00in,bb=40pt 210pt 560pt
615pt]{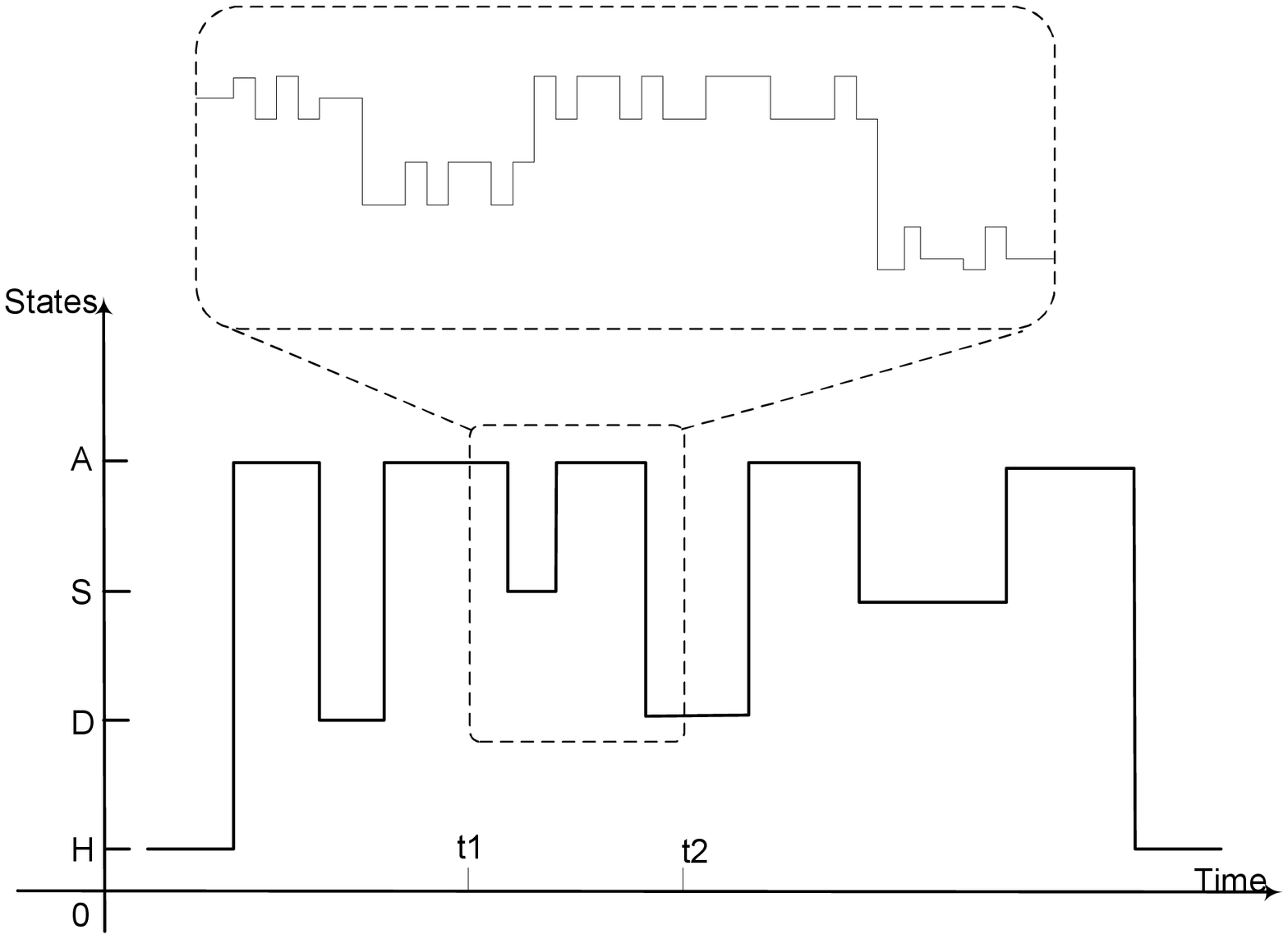}\quad \caption{Continuous-time Markov
chain demonstrated along time  axis $t$. } \label{fig:Multi-time
scale model}
\end{figure}
%Just because of unusual number of tables stacked at end
%\section{Life time of an event and its multi-time scale property}
\bibliography{aps3}% Produces the bibliography via BibTeX.

\end{document}